\documentclass[11pt]{article}

\usepackage[final]{acl}

\usepackage{times}
\usepackage{latexsym}

\usepackage[T1]{fontenc}

\usepackage[utf8]{inputenc}

\usepackage{microtype}

\usepackage{inconsolata}

\usepackage{graphicx}

\usepackage{pifont}
\usepackage{times}
\usepackage{latexsym}
\usepackage{booktabs}
\usepackage{algorithm}
\usepackage{algorithmic}
\usepackage{multicol}
\usepackage{amsfonts}
\usepackage{graphicx}
\usepackage{booktabs}
\usepackage{multirow}
\usepackage{xcolor}
\usepackage{xspace}

\usepackage{makecell} 
\usepackage{array} 
\usepackage{fontawesome} 
\usepackage{xcolor} 
\usepackage{caption} 
\usepackage{amssymb} 
\usepackage{amsmath} 
\usepackage{colortbl}
\usepackage{ulem} 
\usepackage[listings,skins,breakable]{tcolorbox}
\usepackage{pifont}
\usepackage{float} 

\definecolor{aliceblue}{RGB}{255, 238, 241}

\definecolor{babyred}{rgb}{0.85, 0.93, 0.97}

\definecolor{uclablue}{RGB}{159, 195, 224}

\definecolor{uclagold}{RGB}{255, 240, 180}

\definecolor{grayred}{RGB}{205, 186, 207}

\title{KnowPilot: Your Knowledge-Driven Copilot for Domain Tasks}

\author{
    \textbf{Zekun Xi}\thanks{~~Equal contribution.},\hspace{0.5mm}
    \textbf{Yichen Nie$^{*}$},\hspace{0.5mm}
    \textbf{Ziyan Jiang$^{1,*}$},\hspace{0.5mm}
    \textbf{Yujie Bao$^1$},\hspace{0.5mm}\\
    \textbf{Zhenqian Xu$^1$},\hspace{0.5mm}
    \textbf{Zhisong Qiu$^1$},\hspace{0.5mm}
    \textbf{Ziwen Xu$^1$},\hspace{0.5mm} 
    \textbf{Shumin Deng$^{2,}$}\thanks{~~Corresponding Author.},\hspace{0.5mm}\\
    $^1$Zhejiang University\\
    $^2$National University of Singapore\\
    \texttt{\{xizekun2023\}@zju.edu.cn} \\
} 

\begin{document}
\maketitle
\begin{abstract}
Despite the rapid advancement of generative agents, their deployment in real-world industry scenarios often encounters significant challenges due to a lack of domain-specific knowledge. To address this gap, we present KnowPilot: a Domain-Specific Knowledge-Augmented Generative Agent System. KnowPilot is an open-source framework that integrates task-specific priors, explicit knowledge, and experiential knowledge to enhance agent performance in specialized applications. It combines knowledge retrieval from structured repositories with a memory system capable of capturing expert experience through human–AI interaction. Taking domain-specific writing generation as a representative case, KnowPilot enables private deployment, supports injection of task requirements, loads private knowledge bases, and stores tacit expert knowledge as persistent memory. Experimental results demonstrate that KnowPilot achieves superior performance in domain-oriented text generation and is applicable across fields such as medicine, finance and industry\footnote{Video: \url{https://zjunlp.github.io/project/KnowPilot/video}.}.
\end{abstract}

\section{Introduction}

Agent is demonstrating immense potential and value in the current field of artificial intelligence~\cite{yao2023reactsynergizingreasoningacting,nakano2022webgptbrowserassistedquestionansweringhuman}. 
These agents can understand complex natural language instructions, and autonomously plan and execute a wide range of tasks. 
From personal chatbot to automated workflows \cite{qiao2025benchmarkingagenticworkflowgeneration}, the widespread application of agents is profoundly transforming the way we process information and execute tasks.

\begin{figure}[t]
    \centering
    \includegraphics[width=0.95\linewidth]{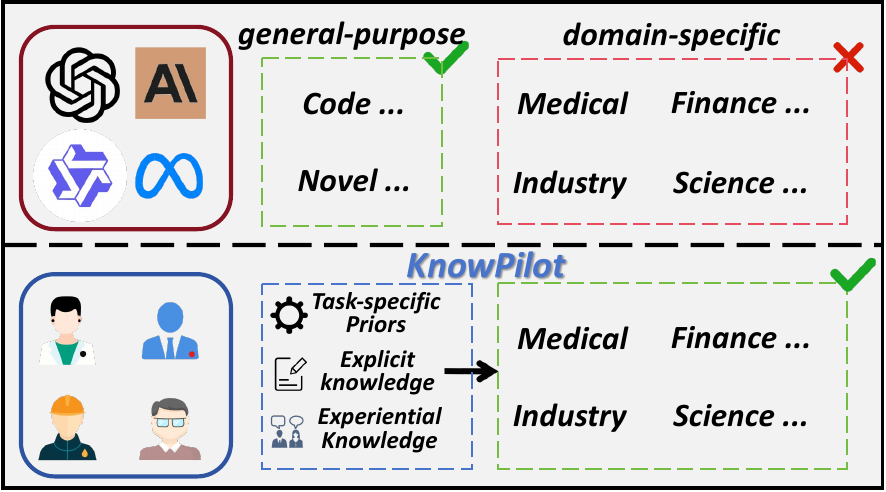}
    \caption{
    Current agents often perform well on generic tasks but struggle to address domain specific problems due to a lack of specialized knowledge. 
    KnowPilot integrates task-specific priors, explicit knowledge, and experiential knowledge to support specific applications.
}
    \label{fig:abs}
\end{figure}

However, current agent systems are largely confined to general applications. 
When it comes to specific domains, such as medical diagnostics, legal consultation \cite{barron2025bridginglegalknowledgeai,cui2024chatlawmultiagentcollaborativelegal}, or financial analysis \cite{li2025hedgeagentsbalancedawaremultiagentfinancial,takayanagi2025generativeaiagentseffective}, existing agents are almost incapable of providing reliable services. 
According to relevant studies, the accuracy of general large language models on specialized tasks, such as medical question answering or legal document analysis, remains substantially lower than their performance on general applications like writing or coding, highlighting a persistent gap in domain specific expertise \cite{guha2023legalbenchcollaborativelybuiltbenchmark,yang2025llmmedqaenhancingmedicalquestion,xi2025omnithinkexpandingknowledgeboundaries}.
The primary cause of this phenomenon is the absence of domain specific knowledge~\cite{song2025injectingdomainspecificknowledgelarge}.
While traditional pre-trained models possess broad general knowledge, they often produce outputs that seem plausible but are factually incorrect when faced with domain specific tasks that require professional judgment and accumulated experience. 
This is unacceptable in high risk professional fields, and there is currently no effective, systematic solution to properly address this knowledge gap.

Injecting domain specific knowledge into agents faces a unique dual challenge \cite{abu2021domain}. 
Firstly, a vast amount of critical domain knowledge exists in private, non-public databases, such as internal corporate documents, reports, and databases. 
Secondly, and more critically, much of the most valuable domain knowledge does not exist in a structured text format but is embodied as \textbf{the tacit knowledge of experts}. 
This experience is deeply embedded in the daily work and decision making of experts and can only be captured and refined through repeated, in-depth interaction and dialogue.

To address these challenges, we propose KnowPilot, a framework designed to systematically endow large language models with domain specific knowledge by combining textual knowledge injection with interactive experience learning, thereby building more professional generative domain specific agents. 
Our core idea is that the model \textbf{should not only inject static, text-based knowledge but also learn dynamic, tacit, practical experiential  knowledge by simulating dialogues with experts, transforming these valuable interactive processes into reusable knowledge assets.}

Experimental results demonstrate that our proposed framework can improve the performance of large language models on domain specific tasks.
KnowPilot enables the agent to go beyond mastering textbook knowledge, allowing it to ``think'' and ``interact'' as a domain expert would, thus providing powerful and dependable support for tackling intricate, high-stakes challenges in real-world scenarios.
We have open-sourced the project code and will provide long-term maintenance.
To conclude, our main contributions are as follows:

\begin{itemize}
    \item We propose the KnowPilot framework, which integrates task-specific priors, explicit knowledge, and experiential knowledge to build knowledge-augmented agents for domain specific tasks.
    \item Our framework enables the accumulation of expert knowledge through natural language interactions, transforming these dialogues into persistent memory that serves as reusable experiential knowledge.
    \item We empirically validate KnowPilot across specific domains, including finance, healthcare, and industry, demonstrating improvements in domain-specific writing tasks.
\end{itemize}
\section{Background}

\begin{figure*}[t]
    \centering
    \includegraphics[width=0.90\linewidth]{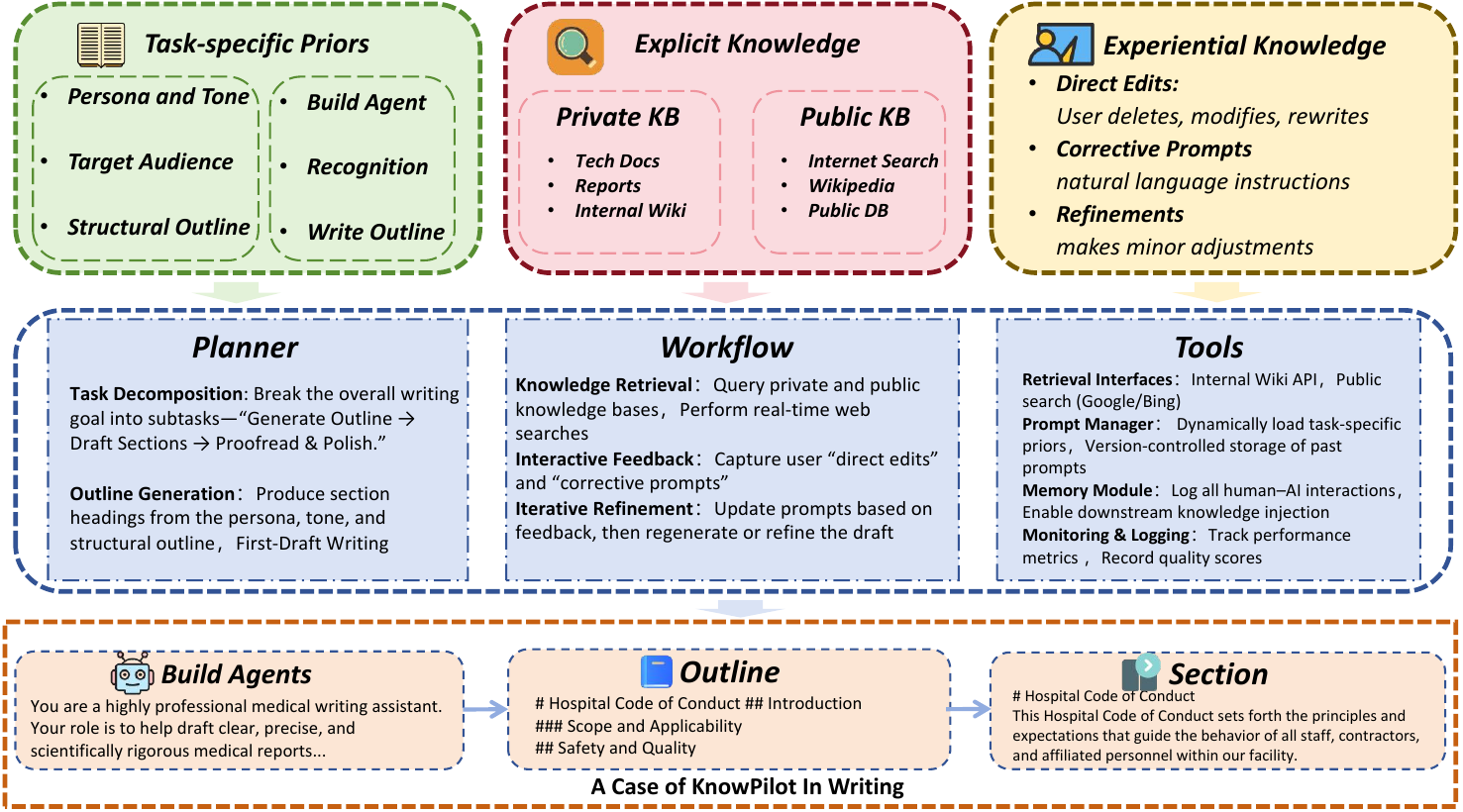}
    \caption{
    The overall architecture of KnowPilot.
    This framework integrates three types of heterogeneous knowledge sources: task-specific priors, experiential knowledge, and explicit knowledge. 
    At the bottom of the figure, we illustrate a case of KnowPilot in domain-specific writing.
}
    \label{fig:main}
\end{figure*}

\paragraph{LLM-based Agents}
Large language models have been widely adopted as the core reasoning engines of agents, providing powerful support for various natural language understanding and generation tasks~\cite{yue2025surveylargelanguagemodel}.
However, due to the static and time-limited nature of the knowledge embedded in pre-trained models, an increasing body of research has focused on augmenting LLMs with external knowledge to address this limitation. 
Typical approaches include retrieval augmentation~\cite{guu2020realmretrievalaugmentedlanguagemodel,borgeaud2022improvinglanguagemodelsretrieving}, which dynamically fetch relevant documents from large corpora during generation to supply factual grounding, as well as integrating knowledge graphs or structured databases to precisely inject entities and relations, thereby reducing factual errors in the output~\cite{peng2024graphretrievalaugmentedgenerationsurvey}. 
Additionally, some studies have explored incorporating tool use directly into the reasoning chain \cite{schick2023toolformer,nakano2022webgptbrowserassistedquestionansweringhuman}, enabling LLMs to proactively initiate API calls, searches, or computations to dynamically supplement the knowledge required for task completion.

While these methods have effectively enhanced breadth and timeliness of knowledge available to agents, they primarily rely on explicitly accessible external resources, such as document repositories, knowledge graphs, or retrieval systems.
They pay relatively little attention to experiential knowledge accumulated over multi-turn interactions, such as learning user preferences, domain-specific styles, or interaction histories. 

\paragraph{Knowledge Augmentation}
Given the inherent limitations of the static knowledge embedded in pre-trained language models, knowledge augmentation has become a central research focus for enhancing the factuality, adaptability, and reasoning capabilities of LLM-based agents~\cite{li2025singleturnsurveymultiturninteractions}.
Early approaches predominantly leveraged retrieval-augmented generation pipelines, where external documents are dynamically retrieved to ground responses in up-to-date information\cite{gao2024modularragtransformingrag}.
There has also been growing interest in integrating multi-modal evidence, such as tables, images, or code snippets, to support complex tasks beyond purely textual reasoning~\cite{li2024benchmarkingmultimodalretrievalaugmented}.

However, most of these efforts focus exclusively on explicit knowledge that can be directly queried or retrieved from external sources. They often overlook another critical dimension of knowledge augmentation: the implicit, experiential knowledge accumulated through sustained interactions.
This includes user-specific writing styles, preferred argument structures, domain heuristics, and even subtle rhetorical choices factors that are difficult to capture through static retrieval alone but are essential for generating content that is not only factually correct but also contextually appropriate and personalized.
In this work, we propose a comprehensive knowledge augmentation strategy that simultaneously integrates task-specific priors, explicit factual knowledge, and experiential knowledge captured from multi-turn interactions.

\section{Design and Implementation}

\subsection{Overview}
\begin{figure}[t]
    \centering
    \includegraphics[width=0.90\linewidth]{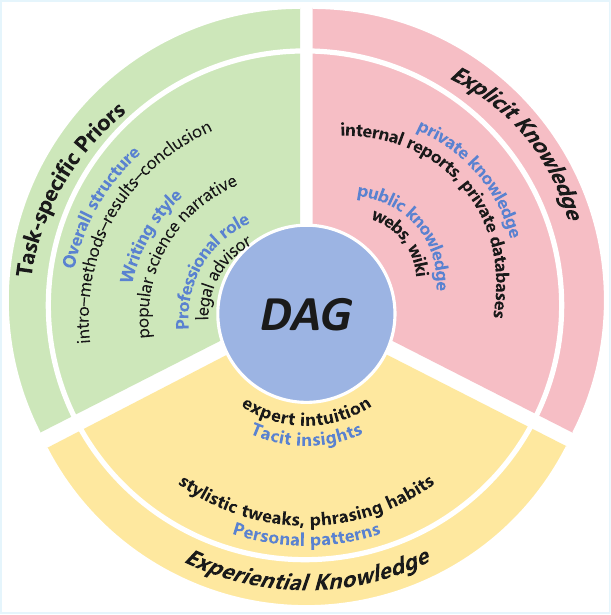}
    \caption{
    Visual illustration of the proposed KnowPilot framework that integrates three heterogeneous knowledge sources for guiding text generation
}
    \label{fig:knowledge}
\end{figure}

The fundamental concept of KnowPilot is the fusion of three heterogeneous knowledge to overcome the limitations of general models. 
As shown in Figure \ref{fig:knowledge}, these include:

\textbf{Task-specific Priors:} Macro level task requirements specified by the user, such as professional role, writing style, and the overall structure.

\textbf{Explicit Knowledge:} Knowledge that can be clearly documented in the text or other forms, which can be retrieved from private or public knowledge bases.

\textbf{Experiential Knowledge:} Tacit insights derived from domain-specific experience, inherently difficult to formalize and often encompassing meanings beyond what words alone can convey.

In KnowPilot, we integrate information to guide content, ensuring that the outputs exhibit factual accuracy, professional style, and logical rigor.

\subsection{Task-specific Priors}
The first step of KnowPilot is to incorporate task-specific priors, setting the tone and overarching structure for the entire generation process to ensure the final output aligns with the user’s high-level objectives.
In terms of implementation, the KnowPilot first queries the user’s requirements and then leverages an LLM to parse these initial instructions, generating a persistent configuration file that solidifies the specifications for persona, style, and related aspects.
Next, guided by the configuration file, the system automatically produces a detailed structural outline of the article.

It should also be noted that the configuration file is not final. 
Users are encouraged to intervene at this stage to review, modify, and refine it, whether by directly editing the sections or by issuing higher-level instructions to reorder chapters or add/remove content.
The operations performed by the user during this interactive refinement process are also captured and stored as Experiential Knowledge.
Furthermore, when parsing the Task-specific Priors, users can optionally choose to incorporate relevant Experiential Knowledge. 
If selected, this Experiential Knowledge is attached to the agent’s prompt, enabling the LLM to interpret the requirements with even greater precision.

By defining the style and structure of the article at the outset, KnowPilot ensures that all subsequent content generation steps remain closely aligned with the user’s core communicative intent.
This effectively prevents directional and structural drift and significantly reduces the cost of later iterative revisions.

\subsection{Explicit Knowledge}
After injecting the task-specific priors, the second step in KnowPilot is to prepare explicit knowledge to provide retrievable and verifiable factual evidence for subsequent content generation~\cite{nonaka1994dynamic}.
Explicit knowledge refers to information that can be clearly documented in text or other forms, which the system can efficiently access to ensure the factual accuracy and professional quality of the generated content.
Specifically, the explicit knowledge base in KnowPilot primarily draws from two sources:

\textit{Private knowledge bases:} KnowPilot can connect to and retrieve information from user-specified internal resources, such as enterprise technical documents, project reports, internal wikis, legal case databases, or market analysis data. In this process, KnowPilot builds a local knowledge base using LangChain based on documents uploaded by the user. Once the LLM returns relevant query keywords, KnowPilot employs Sentence-BERT ~\cite{reimers-gurevych-2019-sentence} to calculate semantic similarity and selects the top 5 most relevant results for retrieval.

\textit{Open-domain sources:} To ensure both the timeliness and breadth of information, the system can also perform real-time searches on the internet or obtain data from large public knowledge bases such as Wikipedia. In practice, we leverage Serper’s Google Search API to search online sources.
By building a rich repository of explicit knowledge at this stage, KnowPilot enables more evidence-based reasoning and articulation in the subsequent generation process.

\subsection{Experiential Knowledge}
The value of an expert lies not merely in the facts they possess, but more importantly in how they apply, organize, and articulate those facts.
During the process of human–AI interaction, users often revise the model’s outputs not because of factual inaccuracies, but due to imprecise wording, weak logical coherence, inappropriate tone, or a failure to reflect domain-recognized best practices.
These very revisions constitute the most valuable experiential data for the system to learn from.
To capture and leverage such data, KnowPilot implements a meticulous recording mechanism designed to track every meaningful user intervention during the final content composition stage.
When users interact with the drafts generated by the model, the system records the following types of behaviors:

\textit{Direct Edits:}
When users delete, modify, or rewrite sentences or paragraphs produced by the model, KnowPilot logs precisely what was changed and how. These operations capture user preferences for particular wordings, enabling more precise expressions in future outputs.

\textit{Corrective Prompts:}
When users provide new natural language instructions to adjust the content, such as “Make this section more objective” or “Add a counter-argument here”, KnowPilot records both the user’s instructions and the before-and-after versions of the text. Such operations help document corrections to the model’s misunderstandings, resulting in more reliable generations over time.

\textit{Refinements:}
After accepting an initial draft, users may ask KnowPilot to make slight adjustments to specific words or phrases to elevate the professional quality of the expression. KnowPilot captures both the original and revised formulations, learning through contrast to better approximate the user’s preferred linguistic style.

All of these records are stored by KnowPilot as reusable experiential knowledge.
When users undertake similar writing tasks in the future, KnowPilot can directly leverage this experiential data to further enhance generation quality and efficiency.

\subsection{Knowledge Fusion}

As shown in Figure \ref{fig:main}, the knowledge fusion process in KnowPilot is a carefully designed pipeline that dynamically integrates task-specific priors, explicit knowledge, and experiential knowledge to systematically guide the generation of high-quality content. 
First, the system loads the user’s Agent configuration file, clarifying high-level goals, stylistic preferences, and structural expectations, while also retrieving experiential knowledge accumulated from past interactions to establish a rich contextual foundation. 
Based on this information, KnowPilot then generates an outline for the article.
At this stage, users are free to edit headings, reorder sections, or add new topics, with the system instantly incorporating these changes to form a robust framework for subsequent writing.
Next, for each section in the outline, the system performs targeted retrievals from private and open-domain knowledge bases, balancing the breadth and specificity of information to ensure that each part is supported by verifiable explicit knowledge. 
Finally, KnowPilot progressively composes the content section by section. Users can adjust or issue new instructions at any time to refine the expression. 
All these interactions are meticulously recorded by the system, consolidating into reusable experiential knowledge that helps future tasks become more efficient and better aligned with user preferences.

The more frequently and deeply users interact with the system, the better the agent can learn their personal preferences, thought patterns, and the domain’s tacit norms.
Thus, every human-AI collaboration transforms from a one-time expenditure of human effort into the creation of a long-term, reusable knowledge asset enabling the agent to truly grow alongside the expert.

\begin{figure*}[t]
    \centering
    \includegraphics[width=0.85\linewidth]{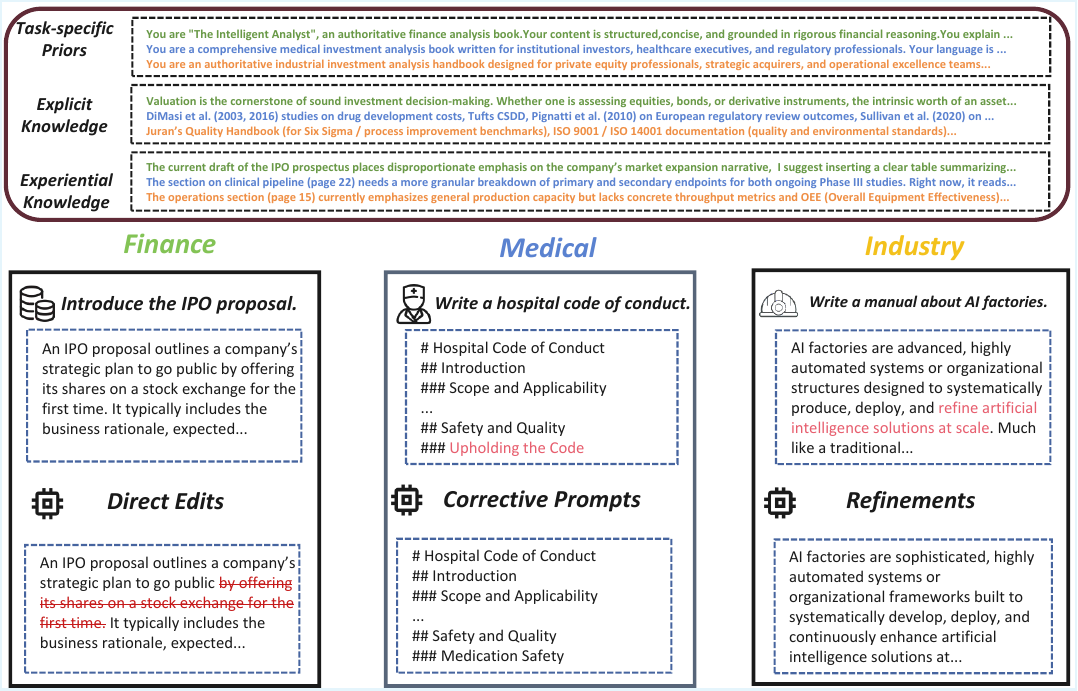}
    \caption{
We present three case studies of KnowPilot in the finance, medical, and industrial domains, illustrating how the KnowPilot integrates  task-specific priors, explicit knowledge and experiential knowledge to more effectively meet user requirements and how users interact with KnowPilot.
}
    \label{fig:case}
\end{figure*}
\subsection{Private deployment}
KnowPilot supports local deployment of large language models via vLLM and Docker, and can operate entirely on private knowledge bases. 
This ensures that all data processing and retrieval stay within secure environments, safeguarding sensitive information and maintaining full control over privacy.

\section{Experiments}
\subsection{Experimental Settings}
We use Wildseek dataset to evaluate the effectiveness of KnowPilot~\cite{jiang2024unknownunknownsengagedhuman}.
Wildseek spans 24 distinct professional domains, providing a comprehensive testbed to assess KnowPilot’s adaptability and generalization capabilities in multi-domain knowledge transfer and collaborative writing scenarios.
We select direct interaction with the LLM and Co-Storm as our baseline methods~\cite{jiang2024unknownunknownsengagedhuman}.
Experiments are conducted using both the open-source large model Qwen3-32B~\cite{qwen3} and the proprietary GPT-4o, ensuring a thorough and fair comparison of performance across different frameworks.

We utilize Prometheus2 to automatically score each method~\cite{kim2024prometheus2opensource}: 
For article quality, we perform quantitative evaluations across completeness, linguistic fluency, and domain relevance, while for outline quality, we provide an overall holistic assessment.
Moreover, as a system that emphasizes interaction, we also focus on measuring interaction time throughout the process, enabling a comprehensive evaluation of KnowPilot’s efficiency in real world applications.

\subsection{Main Results}
\begin{table}[ht]
  \centering
  \label{tab:results}
  \resizebox{0.48\textwidth}{!}{
    \begin{tabular}{l c c c c c}
      \toprule
      \multirow{2}{*}{\makecell[c]{Methods}}      & \multirow{2}{*}{\makecell[c]{Time Score}} & \multirow{2}{*}{\makecell[c]{Outline Score}}
                 & \multicolumn{3}{c}{Article Score} \\
      \cmidrule(lr){4-6}
                 &            &               
                 & Content    & Fluency    & Structure \\
      \midrule
      \multicolumn{6}{c}{\cellcolor{uclablue} \textit{\textbf{GPT-4o}}} \\
      \midrule
      ChatBot    &101.22 &4.19& 3.33 & 3.22 & 3.19\\
      Co-Storm   &312.23 &\textbf{4.41}& 3.42 & \textbf{3.55} & \textbf{3.40}\\
      DAG   &\textbf{100.33} &4.33& \textbf{3.44} & 3.51 & 3.33\\
      \midrule
      \multicolumn{6}{c}{\cellcolor{uclagold} \textit{\textbf{Qwen3-32B}}} \\
      \midrule
      ChatBot    & 111.31 & 3.88 & 3.13 & 2.89 & 2.98\\
      Co-Storm   & 289.01 & 4.01 & \textbf{3.17} & \textbf{3.15} & 3.12\\
      DAG   & \textbf{101.22} & \textbf{4.11} & 3.15 & 3.11 & \textbf{3.23}\\
      \bottomrule
    \end{tabular}
  }
\caption{The results of method across different scoring metrics, Chatbot refers to directly engaging in multi-turn conversations with a large language model}
\end{table}
\paragraph{Domain expertise can be gradually accumulated through sustained dialogue.} 
Through our experiments with KnowPilot, we observed that in many specialized writing scenarios, such as medical reports or legal memoranda, simply accumulating multiturn interaction logs was sufficient to progressively shape language style and content preferences to a level approaching domain expert standards, all without any additional model fine-tuning. This highlights an important insight:
In knowledge-intensive tasks, repeated interactions themselves constitute an implicit form of transfer learning, which over time can partially substitute for resource-intensive domain-specific fine-tuning.

\paragraph{Experiential Knowledge is uniquely valuable: matching performance with reduced effort.}
KnowPilot achieves comparable article quality to Co-Storm while requiring significantly less time, by leveraging user corrections.
Rather than regenerating entire sections from scratch, targeted human refinements quickly steer the content toward domain-appropriate style, logical rigor, and factual precision.
By systematically capturing and reusing these correction patterns, KnowPilot progressively learns to anticipate such adjustments, reducing the need for future interventions even further.
\section{Case Studies}
To further demonstrate the practical utility and workflow of KnowPilot, we present a set of concise case studies derived from typical professional writing scenarios in Figure \ref{fig:case}. 
These examples highlight how KnowPilot captures and leverages multi-turn human–AI interactions to progressively enhance content quality, style, and alignment with user intent. 
Collectively, these cases showcase the core capability of KnowPilot: systematically recording edits, clarifications, and structural adjustments throughout human–AI collaborative writing.
Over time, these accumulated multi-turn interactions are transformed into durable knowledge assets, enabling KnowPilot to reduce redundant iterations, gradually internalize user preferences and stylistic conventions, and ultimately deliver more efficient, personalized, and professional output without the need for costly domain-specific fine-tuning.

\section{Conclusion and Future Work}
In this paper, we introduced KnowPilot, a knowledge-augmented agent system that systematically captures and reuses domain expertise distilled through iterative human–AI collaboration. By fusing pre-defined task specifications, explicit retrievable knowledge, and experiential insights gained from interactive edits, KnowPilot effectively addresses the gap between general-purpose language models and the demands of professional writing scenarios. 
Our experiments across 24 specialized domains demonstrate that KnowPilot not only enhances factual accuracy and stylistic alignment but also progressively internalizes tacit conventions, reducing redundant prompt engineering.
\section{Limitations}

While KnowPilot demonstrates promising results in domain-specific writing generation, several limitations should be acknowledged.

\paragraph{Cold-start problem of Experiential Knowledge.}
The effectiveness of our experiential knowledge mechanism is inherently dependent on the accumulation of sufficient human--AI interaction history. In early-stage deployment, when limited or no prior interactions are available, KnowPilot cannot leverage experiential knowledge to refine its outputs. This cold-start issue may result in suboptimal performance during initial use, particularly in highly specialized domains where tacit expert conventions play a critical role in content quality.

\paragraph{Sensitivity to knowledge base quality.}
KnowPilot relies on both private and open-domain knowledge bases to supply explicit factual grounding. When the underlying documents are outdated, incomplete, or contain inaccuracies, the retrieval pipeline may introduce noise into the generation process. Low-quality or poorly curated knowledge sources can therefore degrade output quality rather than improve it, underscoring the need for careful knowledge base maintenance and curation in practice.

\paragraph{Limited task generalization.}
Our current evaluation is primarily conducted in the context of domain-specific writing generation. Although the proposed framework is conceptually applicable to other knowledge-intensive tasks such as question answering, summarization, and multi-step reasoning, its generalizability to these settings has not yet been empirically validated. The specific design choices optimized for writing workflows, such as outline-driven generation and section-level knowledge fusion, may require adaptation for different task paradigms.

\paragraph{Evaluation methodology constraints.}
The evaluation in this work predominantly relies on automatic metrics computed by Prometheus2. While these metrics provide scalable and reproducible assessments, they may not fully capture nuanced aspects of domain-specific quality such as rhetorical appropriateness, argumentative strength, or adherence to field-specific conventions. Our human evaluation with domain experts, though informative, was limited in scale, and a more comprehensive expert-driven evaluation would strengthen the validity of our findings.

\section{Ethics Statement}

We recognize the importance of ethical considerations in the development and deployment of knowledge-augmented agent systems such as KnowPilot.

\paragraph{Privacy and data protection.}
KnowPilot is designed to connect to private knowledge bases and record detailed user interaction histories, including direct edits, corrective prompts, and refinement operations. Such data may contain sensitive personal or organizational information. To mitigate privacy risks, KnowPilot supports fully local deployment via vLLM and Docker, ensuring that all data processing remains within secure environments. Nevertheless, deployers should implement appropriate access controls, data encryption, and retention policies to safeguard user data in compliance with applicable regulations.

\paragraph{Data usage and consent.}
The experiential knowledge accumulated through human--AI interactions constitutes a form of user-generated data. It is essential that users are clearly informed about what interaction data is being recorded, how it will be stored, and for what purposes it may be reused. Informed consent should be obtained before any interaction data is collected or leveraged for future sessions. We advocate for transparent data governance practices in all deployments of KnowPilot.

\paragraph{Domain bias and fairness.}
Since KnowPilot heavily relies on domain-specific knowledge bases and expert interaction patterns, it may inherit and amplify biases present in the source documents or in the preferences of individual experts. For instance, a medical knowledge base reflecting a particular clinical tradition may lead to outputs that underrepresent alternative therapeutic approaches. Users and deployers should be aware of such potential biases and consider incorporating diverse knowledge sources and multi-expert feedback to promote balanced and fair content generation.

\paragraph{Intended use and misuse prevention.}
KnowPilot is intended to assist domain professionals in producing high-quality written content and should not be used as a substitute for professional judgment in high-stakes decision-making contexts such as clinical diagnosis or legal advice. We encourage responsible use and recommend that all generated outputs undergo expert review before being applied in critical scenarios.

\newpage
\bibliography{custom}




\end{document}